\documentclass[journal]{IEEEtran}
\usepackage[cmex10]{amsmath}
\usepackage{eqparbox}
\usepackage[usenames, dvipsnames]{color}
\usepackage{mathtools,amssymb,bm,mathabx}
\usepackage{amstext}
\usepackage{amssymb}
\usepackage{graphicx}
\usepackage{color}
\usepackage{booktabs}
\usepackage{longtable}
\usepackage{multicol}
\usepackage{amsfonts}
\usepackage{dsfont}
\usepackage{array}
\usepackage{algorithmicx}
\usepackage[ruled]{algorithm}
\usepackage{algpseudocode}
\usepackage{algpascal}
\usepackage{algc}
\usepackage{cases}
\usepackage{pgfplots}
\usepackage{accents}
\usepackage{caption}
\usepackage[footnotesize]{subfigure}
\usepackage{amsthm,xparse}
\DeclareCaptionLabelSeparator{periodspace}{.\quad}
\usepackage{float}
\usepackage{cite}
\usepackage [autostyle, english = american]{csquotes}
\usepackage{hyperref}
\theoremstyle{remark}

\newcommand\ASTART{\bigskip\noindent\begin{minipage}[b]{0.5\linewidth}}
	
	\newcommand\AENDSKIP{\end{minipage}\bigskip}
\newcommand\AEND{\end{minipage}}
\ifCLASSOPTIONcaptionsoff
\usepackage[nomarkers]{endfloat}
\let\MYoriglatexcaption\caption
\renewcommand{\caption}[2][\relax]{\MYoriglatexcaption[#2]{#2}}
\fi
\theoremstyle{plain}
\newtheorem{thm}{\textbf{Theorem}}
\newtheorem{lem}{\textbf{Lemma}}
\newtheorem{prop}{\textbf{Proposition}}

\theoremstyle{definition}

\theoremstyle{remark}
\newtheorem{rem}{\bf Remark}

\newcommand*{\rom}[1]{\expandafter\@slowromancap\romannumeral #1@}
\def\change{black}
\def\chang{black}
\newcommand{\RN}[1]{%
\textup{\uppercase\expandafter{\romannumeral#1}}%
}

\usepackage{standalone}
\graphicspath{{./Figures/}} 

\begin{document}
%
\title{ Distribution-aware Block-sparse Recovery via Convex Optimization}
\author{Sajad~Daei, Farzan~Haddadi, Arash~Amini
	\thanks{S. Daei and F. Haddadi are with the School of Electrical Engineering, Iran University of Science \& Technology. A. Amini is with EE department, Sharif University of Technology.}
}%

\maketitle

\begin{abstract}	
We study the problem of reconstructing a block-sparse signal from compressively sampled measurements. In certain applications, in addition to the inherent block-sparse structure of the signal, some prior information about the block support, i.e. blocks containing non-zero elements, might be available. Although many block-sparse recovery algorithms have been investigated in Bayesian framework, it is still unclear how to incorporate the information about the probability of occurrence into regularization-based block-sparse recovery in an optimal sense. In this work, we bridge between these fields by the aid of a new concept in conic integral geometry. Specifically, we solve a weighted optimization problem when the prior distribution about the block support is available. Moreover, we obtain the unique weights that minimize the expected required number of measurements. Our simulations on both synthetic and real data confirm that these weights considerably decrease the required sample complexity.
\end{abstract}


%
\IEEEpeerreviewmaketitle

\section{Introduction}
 \IEEEPARstart{C}{ompressed} Sensing (CS) has emerged in the past decade as a modern technique for recovering a sparse vector $\bm{x}\in\mathbb{R}^n$ from compressed measurements (see \cite{candes2005decoding,donoho2006most} for more explanations about this field). In this work, we consider signals $\bm{x}\in\mathbb{R}^n$ that have a block-sparse structure, namely their non-zero entries appear in blocks. This property has been referred in the literature as block-sparsity. It is common to use the following optimization problem for recovering the signal $\bm{x}$ from compressive measurements.
  \begin{align}
  \mathsf{P}_{1,2}^{\eta}:~~ &\min_{\bm{z}\in\mathbb{R}^n}\|\bm{z}\|_{1,2}:=\sum_{b=1}^q\|\bm{z}_{\mathcal{V}_b}\|_2,~~ {\rm s.t.} ~\|\bm{A}\bm{z}-\bm{y}\|_2\le \eta,
 \end{align}
 where $\bm{A}\in \mathbb{R}^{m\times n}$ represents a fat measurement matrix with $m\ll n$, $\{\mathcal{V}_b\}_{b=1}^q$ are the default disjoint blocks of size $\{k_b\}_{b=1}^q$ that partition the set $\{1,..., q\}$, $\bm{y}:=\bm{A x}+\bm{e}\in\mathbb{R}^m$ is the observation vector\footnote{Our analysis holds for both real and complex-valued signals and measurements.}, $\bm{e}$ is the noise term which is considered to be i.i.d. Gaussian with variance $\sigma^2$, and $\eta$ is an upper-bound for $\|\bm{e}\|_2$. Most of the earlier literature in block-sparse recovery is focused on the case  of single constraint $\|\bm{A}\bm{z}-\bm{y}\|_2\le \eta$. However, in many applications such as DNA micro-arrays \cite{parvaresh2008recovering,stojnic2008reconstruction}, computational neuroscience \cite{computationalnero} multi-band signal reconstruction, multiple measurement vector (MMV) problem \cite{mishali2008reduce}, and the reconstruction of signals in union of subspaces \cite{eldar2009robust,lu2008theory} \cite{mishali2009blind,mishali2010theory}, there exist additional information (or alternatively additional constraints in $\mathsf{P}_{1,2}^{\eta}$) about the signal of interest. In this work, we explore the benefits of having access to extra information about the distribution of the block support (blocks containing non-zero elements) on the required number of measurements. To this end, we propose the optimization problem
 \begin{align}
 \mathsf{P}_{1,2,\bm{w}}^{\eta}:~~&\min_{\bm{z}\in\mathbb{R}^n}\|\bm{z}\|_{1,2,\bm{w}}:=\sum_{b=1}^q w_b\|\bm{z}_{\mathcal{V}_b}\|_2\nonumber\\
 &\|\bm{y}-\bm{A z}\|_2\le \eta,\nonumber\\
 &\bm{z}\in \mathcal{M}\nonumber
 \end{align} 
 where the quantities $w_b, b=1,..., q$ are some positive scalars, $\bm{w}:=[w_1,..., w_q]^T$ and $\mathcal{M}$ is some predefined model that restricts the feasible set of the solution. Specifically, we consider two new models for prior information that are of practical interest:
\begin{itemize}

	\item Model 1 (\textit{Prior distribution}) : We assume that the prior distribution of the block support is available. Under this setting, there are known probabilities associated with each block index $b\in\{1,..., q\}$. Namely,
	\begin{align}\label{eq.pb}
	\mathds{P}(b\in {\rm bsupp}(\bm{x})) =p_b ~~b=1,..., q,
	\end{align}
	where ${\rm bsupp}(\cdot)$ returns the block support of a vector.
	
	\item Model 2 (\textit{Multiple block support estimates}): We consider {\color{\change} $L$ disjoint sets $\mathcal{P}_i\subset \{1,\dots,q\}$ with $i=1,\dots,L$ that intersect ${\rm bsupp}(\bm{x})$ with the expected accuracy
	\begin{align}\label{eq.alphai}
	\alpha_i:=\frac{\mathds{E}|{\rm bsupp}({\bm x})\cap \mathcal{P}_i|}{|\mathcal{P}_i|}~~i=1,..., L.
	\end{align}}
To each subset $\mathcal{P}_i\subset\{1,..., q\}$, we assign a fixed weight $\lambda_i$. In fact, it holds that 
	\begin{align}\label{eq.w_vs_lambda}
	\bm{w}=\sum_{i=1}^L\lambda_i\bm{1}_{\mathcal{P}_i}{\color{\change}=\bm{D\lambda}\in\mathbb{R}^q,}
	\end{align}
	where $\bm{1}_{C}$ is the indicator function of the set $C$, $\bm{D}:=[\bm{1}_{\mathcal{P}_1},\dots ,\bm{1}_{\mathcal{P}_L}]$, and \color{\change}$\bm{\lambda}:=[\lambda_1, \dots, \lambda_L]^T$.
	
\end{itemize} 

{\color{\change}One of the  applications of Model 1 and 2 is in direction of arrival (DOA) estimation. In this application, Model 1 implies that the probability $p_b$ of having a target in an angle indexed by $b$ is known in advance \cite{mishra2015spectral}. In Model 2, however,  we assume to know the expected number of targets in a range of angles represented by $\mathcal{P}_i$. Such statistics might be available from previous measurements in a dynamic scenario. Obviously, Model 1 imposes more strict conditions as knowing the probabilities for all angles is not easily achievable. In contrast, Model 2 could be applicable as the full angular range could be divided into 3 or 4 intervals, for which we can evaluate the average number of included targets.}
%

In this work, we obtain the weights $\bm{w}^*\in\mathbb{R}^q$ and $\bm{\lambda}^*\in\mathbb{R}^L$ that minimize a threshold $m_0$ describing the expected number of required measurements for Models 1 and 2, respectively. Our approach is to find a suitable upper-bound for $m_0$. The bound is not necessarily tight but leads to closed-form expressions for $\bm{w}\in\mathbb{R}^q$ and $\bm{\lambda}\in\mathbb{R}^L$ in Models 1 and 2, respectively.

\subsection{Related works}
CS in presence of prior information has been studied in different signal models. While a large part of research (see for example \cite{needell2017weighted,khajehnejad2011analyzing,vaswani2010modified,Baraniuk2010,oymak2012recovery,daei2018exploiting}) deals with deterministic signal models, only a few works (see \cite{xu2010compressive,misra2015weighted,diaz2017compressed,fang2015pattern}) have investigated random signal models with Bayesian information. In the deterministic model, the ground-truth signal has intersected with a few sets which called support estimates. The contributing level of each set to the support is available to the experimenter\cite{khajehnejad2011analyzing,needell2017weighted}. This exact situation is investigated in \cite{daei2018exploiting}. They propose a non-uniform model for capturing deterministic prior information. The work \cite{misra2015weighted} considers a probabilistic model where there is a continuous shape function describing the probability of contributing each element to the support. The authors obtain an upper-bound for failure probability of weighted $\ell_1$ minimization. Their approach is based on calculating the internal and external angles of a weighted cross polytope. With a different approach, \cite{diaz2017compressed} has investigated a discrete measure for Bayesian information (a special case of Model 1 with $k=1$). They relate the weights of weighted $\ell_1$ minimization to the discrete probability distribution by minimizing the expected intrinsic volumes of a weighted cone.
{\color{\change}

\subsection{Contributions}
As listed below, we have three main contributions in this work. Besides, our results are also applicable in 
DOA estimation (see Section \ref{section.broaddoa}) and  functional magnetic resonance imaging (fMRI) reconstruction with parallel coils \cite{jung2009k}.
\begin{enumerate}
	\item \textit{Optimally exploiting the block distribution}. In presence of a block distribution, we obtain the optimal weights in $\mathsf{P}_{1,2,\bm{w}}^{\eta}$. This result can be considered as an extension of \cite{diaz2017compressed} to the block-sparse (and joint-sparse\footnote{In this case, the non-zero blocks have common support.}) setting. However, the derivation of the optimal weights in this case is non-trivial and rather challenging. Further, our derivation approach is different from \cite{diaz2017compressed}.
	
	\item \textit{Optimally exploiting multiple estimates.} In presence of multiple block-support estimates, we derive the optimal penalizing coefficients corresponding to each set in weighted $\ell_1$ and $\ell_{1,2}$ minimizations. 
	
	\item \textit{Robustness of optimal weights against inaccurate prior information}. We analytically examine  how close one can get to the optimal weights if the prior information ($p_b$s and $\alpha_i$s) is  inaccurate. This result is important in practical scenarios, as we only have access to the approximations of $p_b$s and $\alpha_i$s. 
	
\end{enumerate}
}
\subsection{Application (Broadband DOA Estimation)}\label{section.broaddoa}
\begin{figure}[t]
	\centering
	\includegraphics[scale=.6]{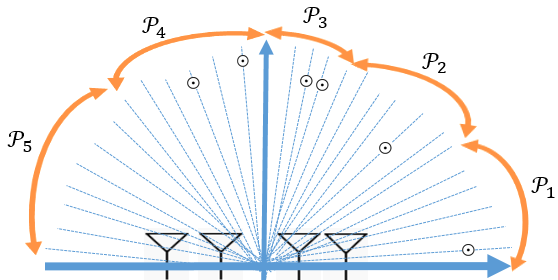}
	\caption{Schematic diagram of DOA estimation of far-field sources. The angular half-space is divided into $q=30$ angular clusters $\{\mathcal{V}_b\}_{b=1}^q$ with equal length. The associated parameters corresponding to Model 2 are $L=5$, $\alpha_1=\frac{1}{6}$, $\alpha_2=\frac{1}{6}$, $\alpha_3=\frac{1}{2}$, $\alpha_4=\frac{1}{3}$ and $\alpha_5=0$.}
	\label{fig.doamodel}
\end{figure}

Suppose that $s$ far-field broadband signals $\{x_i(t,f)\}_{i=1}^s$ in the frequency range $f\in [f_L,f_H]$ incident on an $q$-element uniform linear array (ULA). The received signal in $m$ sensors at time $t$ and $l$-th frequency bin can be expressed as:
\begin{align}
\bm{y}(t,f_l)=\sum_{i=1}^{s}\bm{a}_i(f_l)x_i(t,f_l)+\bm{e}(t,f_l)\in\mathbb{C}^m,
\end{align}
where $\bm{a}_i(f_l):=[1,..., {\rm e}^{-{\rm j}2\pi f_l}\frac{(q-1)d~{\rm sin}(\theta_i)}{c}]^T$ is the steering vector, $c$ is the propagation velocity, $d$ is the inter-sensor spacing and $\bm{e}(t,f_l)$ is a Gaussian noise term with variance $\sigma^2$. In practice, one has to take several snapshots $\{\bm{y}(t,f_l)\}_{t=1}^k$. This temporal redundancy is crucial in practice since the array size is limited due to physical constraints \cite{yang2016exact,hyder2010direction}. Consequently, one may write
\begin{align}
&\bm{Y}(f_l):=\left[\bm{y}(1,f_l),..., \bm{y}(k,f_l)\right]=\nonumber\\
&\bm{A}(f_l)\bm{X}(f_l)+\bm{E}(f_l)\in\mathbb{C}^{m\times k},
\end{align}
where $\bm{X}(f_l)=\left[\bm{x}(1,f_l),..., \bm{x}({\color{\change}k},f_l) \right]\in\mathbb{C}^{q\times k}$, $\bm{A}(f_l)=\left[\bm{a}_1(f_l),..., \bm{a}_q(f_l)\right]\in\mathbb{C}^{m\times q}$ and $\bm{E}(f_l)\in\mathbb{C}^{m\times k}$ is defined similar to {\color{\change} $\bm{Y}(f_l)$} . If the sources is time-invariant over the period of snapshotting, then for all $t=1,..., k$ the non-zero dominant peaks in $\bm{x}(t,f_l)$ occur at the same locations corresponding to the ground-truth DOAs. Hence, DOA estimation can be cast as recovering a joint sparse signal $\bm{X}(f_l)$ from $\bm{Y}(f_l)$. In addition, it is realistic for a radar engineer to know the probability of appearing the ground-truth DOAs in some angular bands \cite{mishra2015spectral} (see Figure \ref{fig.doamodel} for a schematic model of this scenario).

\textit{Notation}. Throughout, scalars are denoted by lowercase letters, vectors by lowercase boldface letters, and matrices by uppercase boldface letters. The $i$th element of a vector $\bm{x}$ is shown either by ${x}(i)$ or $x_i$. $\mathcal{C}^\circ$ denotes the polar of a cone $\mathcal{C}$. We show sets (e.g. $\mathcal{B}$) by calligraphic uppercase letters. {\color{\change} We show the set $\{1, \dots, n\}$ by $[n]$.} ${\mathcal{\bar{B}}}$ is used to represent the complement $[n]\setminus\mathcal{
	B}$ of a set $\mathcal{B}\subset [n]$. 

\section{Main results}\label{section.mainresult}
 In the following propositions, we obtain closed-form solutions for optimal weights in case that $\bm{x}$ satisfies Model 1 and 2. The proofs are provided in Appendix \ref{proof.prop1and2}.
 
\begin{prop}\label{prop.dist}
Let $\bm{x}\in\mathbb{C}^n$ satisfy Model 1 with parameter $\bm{p}=[p_1,..., p_q]^T$. Then, the optimal weights $\bm{w}^*=[w_1^*,...,w_q^*]^T$ in $\mathsf{P}_{1,2,\bm{w}}$ are obtained by solving the following equations simultaneously:
\begin{align}\label{eq.optimalweights_dist}
&\frac{p_b}{1-p_b}w^*_b=\frac{1}{2^{\frac{k_b}{2}-1}\Gamma(\frac{k_b}{2})}\int_{w^*_b}^{\infty}(u-w^*_b)u^{k_b-1}{\rm{e}}^{-\frac{u^2}{2}}{\rm{d}}u,\nonumber\\
&b=1,..., q.
\end{align}
\end{prop}

{\color{\change}
\begin{rem}(Prior work)
The special case $k_b=1$ ($\bm{x}$ is sparse instead of block-sparse) reduces (\ref{eq.optimalweights_dist}) to the weighted $\ell_1$ minimization which is studied in \cite{diaz2017compressed}. Therefore, Proposition \ref{prop.dist} generalizes the results of \cite{diaz2017compressed} to the block-sparse case. However, our approach to reach this generalized result is different from and somewhat simpler than \cite{diaz2017compressed}.
\end{rem}
}

\begin{prop}\label{prop.mutiple_estimate}
Let $\bm{x}\in\mathbb{C}^n$ be decomposed into $q$ blocks $\mathcal{V}_b$ of equal size $k$. Assume that there exist $L$ independent estimates $\{\mathcal{P}_i\}_{i=1}^L$ of ${\rm bsupp}(\bm{x})$ with parameter $\bm{\alpha}=[\alpha_1,..., \alpha_L]^T$. Then, the optimal weights $\bm{\lambda}^*=[\lambda_1^*,..., \lambda_L^*]^T$ in Model 2 are obtained by solving
\begin{align}\label{eq.optimalweights_estimate}
&\frac{\alpha_i}{1-\alpha_i}\lambda^*_i=\frac{1}{2^{\frac{k}{2}-1}\Gamma(\frac{k}{2})}\int_{\lambda^*_i}^{\infty}(u-\lambda^*_i)u^{k-1}{\rm{e}}^{-\frac{u^2}{2}}{\rm{d}}u,\nonumber\\
&i=1,..., L.
\end{align}
\end{prop}

{\color{\chang}
\begin{rem}
The optimal weights $w_b^*$s and $\lambda_i^*$s in Propositions \ref{prop.dist} and \ref{prop.mutiple_estimate} are respectively obtained by minimizing an upper-bound of the expected number of required measurements in problems $\mathsf{P}_{1,2,\bm{w}}^{0}$ and $\mathsf{P}_{1,2,\bm{D}\bm{\lambda}}^{0}$. We numerically observe that the upper-bound is tight for non-uniform distributions of $\bm{x}$, but the exact identification of such distributions is beyond the scope of this work. Further, the system of equations in \eqref{eq.optimalweights_dist} and \eqref{eq.optimalweights_estimate} are solved using the $\rm fzero$ function in MATLAB.
\end{rem}}

\begin{figure}[t]
	\hspace{-.4cm}
	\includegraphics[scale=.2]{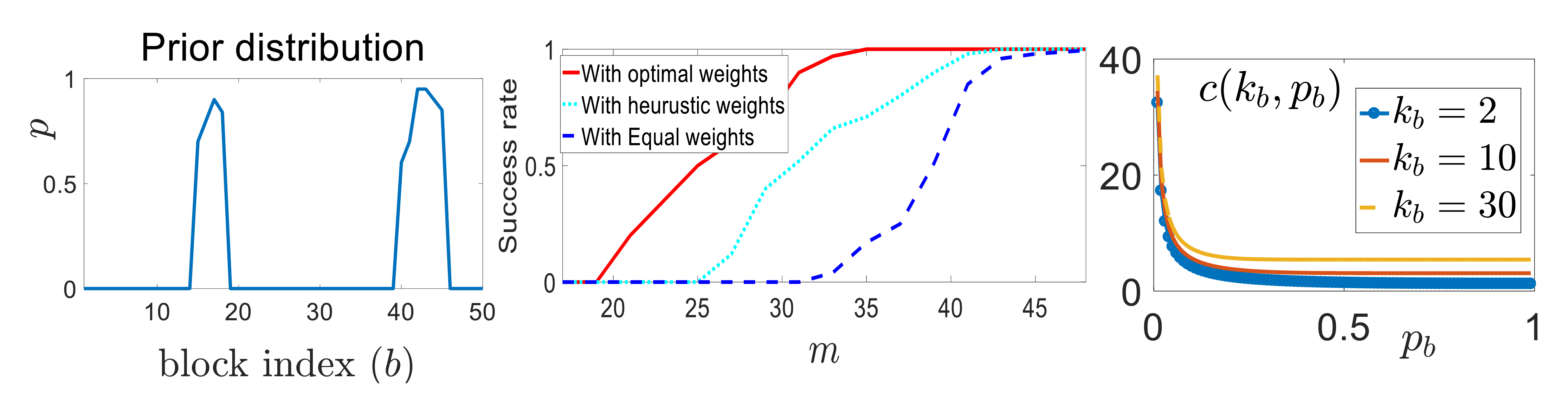}
	\caption{The middle image compares the success rate of $\mathsf{P}_{1,2,\bm{w}^*}^{0}$ with that in $\mathsf{P}_{1,2}^0$ when there exists a prior distribution $\bm{p}\in\mathbb{R}^q$ (depicted in the left image) about the block indexes $\{1,..., 50\}$. The optimal weight $\bm{w}^*$ is obtained by (\ref{eq.optimalweights_dist}). The right image represents $c(k_b,p_b)$ in Theorem 1\ref{prop.robustness} for different $k_b$'s.}
	\label{fig.succ+prior+robust}
\end{figure}

In applications, it is of important practical value to know how the inaccuracies of $p_b$ and $\alpha_i$ in Model 1 and 2, respectively, affect the optimal weights. The following theorem is about this challenge.
\begin{thm}\label{prop.robustness}
	Assume that $\bm{p}$ and $\bm{p}'$ be the true and approximate estimate of ${\rm{bsupp}}(\bm{x})$, respectively. Let $\bm{w}$ and $\bm{w}'$ be the optimal weights corresponding to $\bm{p}$, and $\bm{p}'$. Then, there exists a constant $c(k_b,p_b)$ such that
	\begin{align}
	&|w_b-w_b'|\le c(k_b,p_b)|p_b-p_b'| ~~b=1,..., q\nonumber\\
	&c(k_b,p_b):=\nonumber\\
	&\frac{\Bigg(\sqrt{2}h(p_b)\big(\Gamma(\frac{k_b}{2})-\gamma(\frac{k_b}{2},\frac{h(p_b)^2}{2})\big)+2\gamma(\frac{k_b}{2},\frac{h(p_b)^2}{2})\Bigg)^2}{2\sqrt{2}\Gamma(\frac{k_b}{2})\gamma(\frac{k_b}{2},\frac{h(p_b)^2}{2})}
	\end{align}
	where $\gamma(a,z):=\int_{z}^{\infty}u^{a-1}{\rm{e}}^{-u}{\rm{d}}u$ is incomplete gamma function and $h(\cdot):p_b\rightarrow w_b$ is the nonlinear function in (\ref{eq.optimalweights_dist}).
\end{thm}
From the above theorem and the right image of Figure \ref{fig.succ+prior+robust}, one can infer that the method of obtaining $w_b^*$ and $\lambda_i^*$ is robust to slight changes of $p_b$ and $\alpha_i$ as long as they are greater than {\color{\change} approximately} $\frac{1}{10}$.

\begin{figure}[t]
	\hspace{-0.44cm}
	\includegraphics[scale=.2]{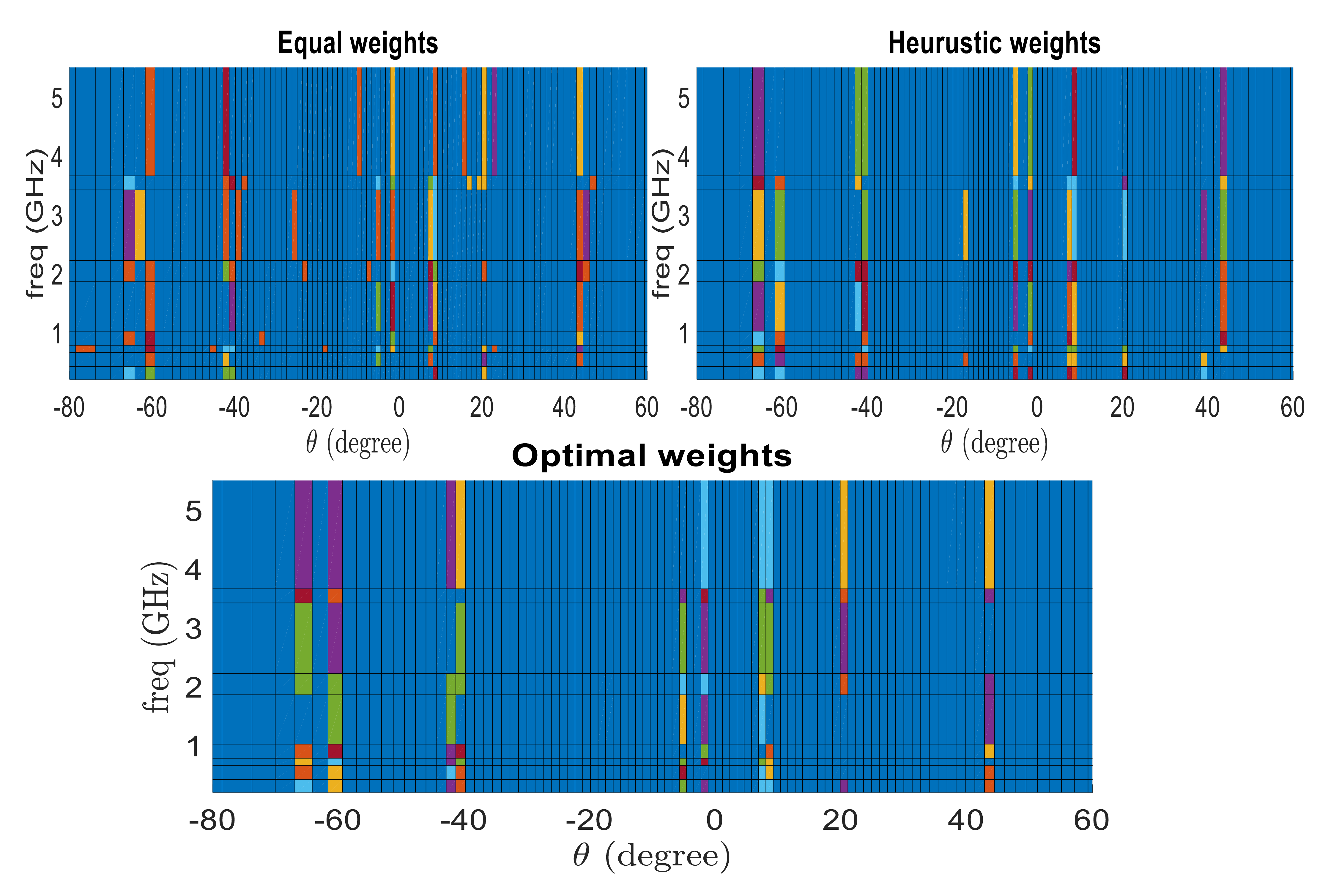}
	\caption{Broadband DOA estimation for $10$ sources {\color{\chang}(located at the angles $-66.9^\circ$, $-61.64^\circ$, $-42.84^\circ$, $-41.3^\circ$, $-5.74^\circ$, $-2.3^\circ$, $6.89^\circ$, $8.05^\circ$, $19.88^\circ$, $42.84^\circ$)} using an $m=15$ sensor ULA. The associated parameters are $k=10$, $q=100$, $d=5$, $c=3\times 10^8$, and $\sigma=1$. {\color{\change}The expected accuracy corresponding to three sets $\{\mathcal{P}_i\}_{i=1}^3$ are $\alpha_1=\frac{4}{5}$, $\alpha_2=\frac{2}{3}$, and $\alpha_3=0$. The top left, top right and bottom images correspond to the recovery using  $\mathsf{P}_{1,2,\bm{D}\bm{\lambda}}^{\eta}$, when equal, heuristic and optimal weights $\bm{\lambda}$ (calculated using (\ref{eq.optimalweights_estimate})) are applied, respectively.}}
	\label{fig.doabroad}
\end{figure}

\vspace{-.5cm}

\section{Simulations}\label{section.simulation}
In the first experiment, we construct a random block-sparse $\bm{x}\in\mathbb{R}^{250}$, whose building blocks $\mathcal{V}_b$ have equal size $k=5$. The probability of activating each block is taken from the vector $\bm{p}$ depicted in the left image of Figure \ref{fig.succ+prior+robust} . Then, this signal is observed through a measurement matrix $\bm{A}\in\mathbb{R}^{m\times 250}$, of which the elements are drawn from i.i.d. standard normal distribution. We obtain the optimal weights $\bm{w}^*$ corresponding to $\bm{p}$ by solving (\ref{eq.optimalweights_dist}). {\color{\change}We also examine the heuristic weights $w_b=\tfrac{1}{p_b+\epsilon}$.} In the middle image of Figure \ref{fig.succ+prior+robust}, we plot the success rate as a function of $m$. For each $m$, we average over $100$ realizations of $\bm{A}$ and $\bm{x}$.  {\color{\change}As expected,  $\mathsf{P}_{1,2,\bm{w}}^{0}$ requires fewer measurements for successful recovery with optimal weights $\bm{w}^*$ rather than the other two options (heuristic or equal weights). In turn, heuristic weights are also superior to the equal weights.
}


In the second experiment, we test DOA estimation using broadband signals (see Subsection \ref{section.broaddoa} and Figure \ref{fig.doamodel}). The angular half-space $-[90^\circ,90^\circ)$ is divided into $q=100$ angular grids. For each frequency bin $f_l\in[0,5]$ GHz, we take $k=10$ snapshots. We assume that there exist three 
{\color{\change}($L=3$) sets $\mathcal{P}_i$ with expected accuracies} 
$\alpha_1=\frac{4}{5}$, $\alpha_2=\frac{2}{3}$, and $\alpha_3=0$. Also, we use $m=15$ sensors for recovering $s=10$ ground-truth sources {\color{\chang} (located at the angles $-66.9^\circ$, $-61.64^\circ$, $-42.84^\circ$, $-41.3^\circ$, $-5.74^\circ$, $-2.3^\circ$, $6.89^\circ$, $8.05^\circ$, $19.88^\circ$, $42.84^\circ$)} and implement the optimization problem 
{\color{\change} $\mathsf{P}_{1,2,\bm{D \lambda}}^{\eta}$ when $\bm{\lambda}$ is chosen optimally (i.e. $\lambda_i^*$'s are obtained using (\ref{eq.optimalweights_estimate})), heuristically (i.e. $\lambda_i=\tfrac{1}{\alpha_i+\epsilon}$) and equally}. 
As it turns out from Figure \ref{fig.doabroad}, {\color{\change} while $\mathsf{P}_{1,2,\bm{D \lambda}}^{\eta}$ with equal and heuristic weights detects [respectively many and a few] non-existing sources with spurious DOAs, $\mathsf{P}_{1,2,\bm{D}\bm{\lambda}^*}^{\eta}$ locates the ground-truth sources correctly}. This in turn suggests that our optimal weighting strategy considerably decreases the required number of sensors.
\vspace{0cm}
\appendix
\subsection{Preliminaries}
In this section, we introduce two concepts from conic integral geometry that are used in our analysis.

\textit{Descent cone}: Let $\bm{x}$ be a vector {\color{\change} with a special low-dimensional structure (e.g. block-sparsity). Assume that $f$ is a convex function that promotes this structure}. Then, the set of decent directions forms a convex set defined by:
\begin{align}\label{eq.descent cone}
\mathcal{D}(f,\bm{x})=\bigcup_{t\ge0}\{\bm{z}\in\mathbb{C}^n: f(\bm{x}+t\bm{z})\le f(\bm{x})\}\cdot
\end{align}
\textit{Statistical dimension}: {\color{\change} Statistical dimension is intuitively a measure for the size of a cone.} It is shown in \cite{amelunxen2013living} that statistical dimension of the above decent cone defined by
\begin{align}
\delta(\mathcal{D}(f,\bm{x})):=\mathds{E}_{\bm{g}}{\rm dist}^2(\bm{g},\mathcal{D}(f,\bm{x})^\circ),
\end{align}
specifies the required number of measurements (i.e. $m$) that
\begin{align}
\mathsf{P}_f: ~\min_{\bm{z}\in\mathbb{C}^n}f(\bm{z}) ~~~{\rm s.t.} ~~\bm{y}_{m\times 1}=\bm{A z}
\end{align}
needs for perfect recovery. {\color{\change}Here, $\bm{g}\in\mathbb{R}^n$ is a vector with i.i.d. standard normal distribution.} We define the expected number of measurements needed for $\mathsf{P}_{1,2,\bm{w}}$ as
\begin{align}
\overline{m}:=\mathds{E}_{\bm{x}}\delta(\mathcal{D}(\|\cdot\|_{1,2,\bm{w}},\bm{x})).
\end{align}
Then, we call the weights that minimize $\overline{m}$ optimal in sense of expected sample complexity.

\subsection{Proof of Propositions \ref{prop.dist} and \ref{prop.mutiple_estimate}}\label{proof.prop1and2}
\begin{proof}Since $\delta(\mathcal{D}(\|\cdot\|_{1,2,\bm{w}},\bm{x}))$ is {\color{\change}upper-bounded by an expression} that only depends on ${\rm{bsupp}}(\bm{x})$ and $\bm{w}$ (see Lemma \ref{lem.upperbound}), we show {\color{\change}the corresponding upper-bound} by $\delta(\mathcal{D}(\mathcal{B},\bm{w}))$. It holds that:
	\begin{align}\label{eq.mhatqsw}
	&\overline{m}\le\mathds{E}_{\bm{x}}\delta(\mathcal{D}({\rm bsupp}(\bm{x}),\bm{w})){\color{\change}:=}
\mathds{E}_{\bm{x}}\delta(\mathcal{D}(\mathcal{B},\bm{w}))
	\end{align}
	We proceed by using a {\color{\change} closed-form expression} for $\delta(\mathcal{D}(\mathcal{B},\bm{w}))$ a special case of which is obtained in \cite[Lemma {\color{\change}2}]{daei2018exploiting}.
	\begin{lem}\label{lem.upperbound}
		The statistical dimension of descent cone of any vector $\bm{x}_{n\times 1}$ with ${\rm bsupp}(\bm{x}):=\mathcal{B}$ satisfies:
	\begin{align}
	&\delta(\mathcal{D}(\|\cdot\|_{1,2,\bm{w}},\bm{x}))\le\delta(\mathcal{D}(\mathcal{B},\bm{w})){\color{\change}:=}\nonumber\\
	& \inf_{t\ge 0}\Big(\sum_{b\in \mathcal{B}}\big(k_b+(tw_b)^2\big)+\tfrac{\sum_{b\in {\mathcal{\overline{B}}}}\phi_B(tw_b,k_b)}{2^{\tfrac{k_b}{2}-1}\Gamma(\tfrac{k_b}{2})}\Big),\nonumber\\
	&\text{with}~~\phi_B(z,k):=\int_{z}^{\infty}(u-z)^2u^{k-1}\exp(-\frac{u^2}{2}){\rm{d}}u.\nonumber
	\end{align}
	Moreover, the minimum is achieved at a unique $t\ge 0$. The inequality is in fact equality in the asymptotic case ($q\rightarrow \infty$)\cite[Proposition 3]{daei2018exploiting}. 
	\end{lem}
Thus, by the aid of Lemma \ref{lem.upperbound}, it holds that
{\color{\change}
\begin{align}
&\overline{m}{\color{\change}\stackrel{(\RN{1})}{\le}}\mathds{E}\inf_{t\ge 0}\sum_{b=1}^q \Big(\big(k_b+(tw_b)^2\big)1_{b\in\mathcal{B}}+\tfrac{\phi_B(tw_b,k_b)1_{b\in{\overline{\mathcal{B}}}}}{2^{\frac{k_b}{2}-1}\Gamma(\frac{k_b}{2})}\Big){\color{\change}\stackrel{(\RN{2})}{\le}} \nonumber\\
&\inf_{t\ge 0}\sum_{b=1}^q
\Big(\big(k_b+(tw_b)^2\big)\underbrace{\mathds{E}\left[1_{b\in\mathcal{B}}\right]}_{p_b}+\tfrac{\phi_B(tw_b,k_b)\overbrace{\mathds{E}\left[1_{b\in{\overline{\mathcal{B}}}}\right]}^{1-p_b}}{2^{\frac{k_b}{2}-1}\Gamma(\frac{k_b}{2})}\Big)
\end{align}
where in the inequality $(\RN{1})$, the notation $1_{\mathcal{E}}$ signifies the indicator function of an event $\mathcal{E}$. The inequality $(\RN{2})$ results from the Jensen inequality for concave functions.} Therefore, we have:
{\color{\change}
\begin{align}\label{eq.m_bar1}
\overline{m}\le\inf_{t\ge 0}\sum_{b=1}^{q}\Bigg[p_b(k_b+t^2w_b^2)+\frac{(1-p_b)}{2^{\frac{k_b}{2}-1}\Gamma(\frac{k_b}{2})}\phi_B(tw_b,k_b)\Bigg].
\end{align}
Moreover, by partitioning the set $[q]$ into the sets $\mathcal{P}_i$s, one may write:
\begin{align}
&\overline{m}\le\inf_{t\ge 0}\sum_{i=1}^L\sum_{b\in\mathcal{P}_i}\Big(\big(k_b+(tw_b)^2\big)p_b+\tfrac{(1-p_b)\phi_B(tw_b,k_b)}{2^{\frac{k_b}{2}-1}\Gamma(\frac{k_b}{2})}\Big)=\nonumber\\
&\sum_{b=1}^q k_b p_b+\inf_{t\ge 0}\sum_{i=1}^L\sum_{b\in\mathcal{P}_i}\left[ t^2\lambda_i^2p_b+\frac{(1-p_b)}{2^{\frac{k_b}{2}-1}\Gamma(\frac{k_b}{2})}\phi_B(t\lambda_i,k_b) \right]
\end{align}
where we used the relation (\ref{eq.w_vs_lambda}) in the last equality. By further assumption $k_b=k ~\forall b$, it holds that:
\begin{align}
&\overline{m}\le \sum_{b=1}^{q} k_b p_b+\inf_{t\ge 0}\sum_{i=1}^{L}\Big[ t^2\lambda_i^2\sum_{b\in\mathcal{P}_i}p_b+\tfrac{\phi_B(t\lambda_i,k)\sum_{b\in\mathcal{P}_i}(1-p_b)}{2^{\tfrac{k}{2}-1}\Gamma(\tfrac{k}{2})} \Big].
\end{align}
Since $\alpha_i:=\tfrac{\mathds{E}|{\rm bsupp}(\bm{x})\cap \mathcal{P}_i|}{|\mathcal{P}_i|}=\tfrac{\sum_{b\in\mathcal{P}_i}p_b}{|\mathcal{P}_i|}$ and $1-\alpha_i=\tfrac{\mathds{E}|\overline{{\rm bsupp}(\bm{x})}\cap \mathcal{P}_i|}{|\mathcal{P}_i|}=\tfrac{\sum_{b\in\mathcal{P}_i}(1-p_b)}{|\mathcal{P}_i|}$, we have:
\begin{align}\label{eq.mbar2}
&\overline{m}\le \sum_{b=1}^{q} k_b p_b+\inf_{t\ge 0}\sum_{i=1}^{L}\Bigg[ t^2\lambda_i^2\alpha_i+\frac{1-\alpha_i}{2^{\frac{k}{2}-1}\Gamma(\frac{k}{2})}\phi_B(t\lambda_i,k)\Bigg].
\end{align}}
{\color{\chang} As multiplication with a positive scalar $t$ keeps the optimal choices of $w_b$ and $\lambda_i$ in $\mathsf{P}_{1,2,\bm{w}^*}$ and $\mathsf{P}_{1,2,\bm{D}\bm{\lambda}^*}$ untouched}, by minimizing the expressions in brackets in (\ref{eq.m_bar1}) and (\ref{eq.mbar2}) with respect to $w_b$ and $\lambda_i$ {\color{\chang}(the second derivatives of these expressions are always positive with respect to $w_b$ and $\lambda_i$ and hence they are strictly convex functions)}, one can get to the expressions (\ref{eq.optimalweights_dist}) and (\ref{eq.optimalweights_estimate}).
\end{proof}
\ifCLASSOPTIONcaptionsoff
  \newpage
\fi

\bibliographystyle{ieeetr}
\bibliography{mypaperbibe1}
\end{document}